\documentclass[
 reprint,
 amsmath,amssymb,
 aps,
 prr,
 superscriptaddress,
]{revtex4-2}

\usepackage[utf8]{inputenc}
\usepackage[T1]{fontenc}
\usepackage{graphicx}
\usepackage{hyperref}
\usepackage{gensymb}
\usepackage{xcolor}
\usepackage{placeins}
\usepackage{siunitx}
\usepackage{gensymb}
\usepackage{textcomp}
\usepackage[version=4]{mhchem}

\newcommand{\iso}[2]{\ensuremath{^{#2}\mathrm{#1}}}
\usepackage{xcolor}

\begin{document}

\preprint{APS/123-QED}

\title{Laser cooling and trapping of \texorpdfstring{$^{224}$Ra$^+$}{Ra-224 ions}}

\author{M. Fan}
\email{mingyufan212@gmail.com}
\author{Roy A. Ready}
\author{H. Li}
\author{S. Kofford}
\author{R. Kwapisz}
\author{C. A. Holliman}
\author{M. S. Ladabaum}
\affiliation{Department of Physics, University of California, Santa Barbara, California 93106, USA}
\author{A. N. Gaiser}
\affiliation{Department of Chemistry, Michigan State University, East Lansing, Michigan 48824, USA}
\affiliation{Facility for Rare Isotope Beams, Michigan State University, East Lansing, Michigan 48824, USA}
\author{J. R. Griswold}
\affiliation{Radioisotope Science and Technology Division, Oak Ridge National Laboratory, Oak Ridge, Tennessee 37830, USA}
\author{A. M. Jayich}
\affiliation{Department of Physics, University of California, Santa Barbara, California 93106, USA}

\date{\today}

\begin{abstract}
We report laser cooling and trapping of \iso{Ra}{224}$^+$ ions.  This was realized via two-step photoionization loading of radium into an ion trap.  A robust source for \iso{Ra}{224} atoms, which have a 3.6-day half-life, was realized with an effusive oven containing \iso{Th}{228}, which has a 1.9-yr half-life, which continuously generates \iso{Ra}{224} via its $\alpha$-decay.  We characterized the efficacy of this source and found that after depleting built-up radium the thorium decay provides a continuous source of radium atoms suitable for ion trapping. The vacuum system has been sealed for more than 6 months and continues to trap ions on demand.  We also report a measurement of the \iso{Ra}{224} \mbox{$7s^2\ ^1$S$_0 \rightarrow 7s7p\ ^1$P$_1$} transition frequency: \SI{621043830}{}$\pm60$ MHz, which is helpful for efficient photoionization.  With this measurement and previous isotope shift measurements we find that the frequency of the same transition in \iso{Ra}{226} is \SI{621037830}{}$\pm60$ MHz, which disagrees with the most precise measurement, \SI{621038489}{}$\pm15$ MHz, which is used for the recommended value in the National Institute of Standards and Technology Atomic Spectra Database.
\end{abstract}

\maketitle

\section{introduction}

Radium, the heaviest alkaline earth element, has favorable electronic properties for laser cooling and trapping in both neutral and singly ionized forms~\cite{Guest2007,Fan2019}. Ra$^+$ has a narrow-linewidth electric quadrupole (E2) transition, which is advantageous for trapped-ion optical clocks~\cite{Sahoo2007, Versolato2011b, Holliman2022}. The Ra$^+$ ion clock operates with wavelengths that are compatible with integrated photonic technologies, which makes Ra$^+$ appealing for a transportable optical clock. Certain isotopes of radium, such as \iso{Ra}{225} ($I=1/2$), have an octupole deformed nucleus which when paired with their nuclear spin enhances sensitivity to time-reversal symmetry violations~\cite{Parker2015}.  A challenge with radium is that there are no stable isotopes.  The longest lived, \iso{Ra}{226} ($I=0$), has a 1600-yr half-life, while \iso{Ra}{225} has only a 15-day half-life.  For all radium isotopes, radioactivity limits their use to small quantities.

Previous atomic and molecular experiments used a variety of mechanisms to work with various radium isotopes:   Spectroscopy of trapped \iso{Ra}{209-214}$^+$ ions was performed at the TRI$\mu$P nuclear facility, where radium atoms were produced when a lead beam impinged on a carbon target~\cite{Giri2011a}. An optical atomic clock was demonstrated with a \iso{Ra}{226}$^+$ ion, where the Ra$^+$ was produced via laser ablation of a radium chloride target in a vacuum system~\cite{Holliman2022}. The atomic electric dipole moment of neutral \iso{Ra}{225} was measured in an optical dipole trap, where \iso{Ra}{225} was directly loaded into an effusive oven and heated out for laser cooling and trapping~\cite{Parker2015}.  Radium isotopes were produced at the European Organization for nuclear Research (CERN) by using 1.4-GeV protons to irradiate a uranium carbide target which was then heated to release radium atoms to form RaF for spectroscopy \cite{GarciaRuiz2020}.  When working with all but \iso{Ra}{226}, these techniques require specialized facilities and or breaking vacuum on timescales incommensurate with typical atomic and molecular experiments. Fortunately, thorium may be used as a generator to continuously produce \emph{in vacuo} \iso{Ra}{224}, \iso{Ra}{226}, and the desirable \iso{Ra}{225} isotope, relieving the need for nuclear facilities or opening vacuum systems.  This method was used for spectroscopy of neutral \iso{Ra}{225} from an effusive oven~\cite{Santra2013,Santra2014}.

Thorium has a vapor pressure that is more than $10^{12}$ times lower than that of radium~\cite{haynes2015}; therefore when an oven is heated, it will produce a radium beam while a negligible quantity of thorium will leave the oven. An oven based on this mechanism should provide radium for several thorium half-lives. We use an effusive oven based on this mechanism to realize the first photoionization loading of radium ions into an ion trap and the first laser cooling of \iso{Ra}{224}$^+$ ions.  Radium-224, half-life 3.6~days~\cite{Bergeron2021}, is continuously produced by the $\alpha$-decay of \iso{Th}{228}, half-life 1.9~yr, in the oven's crucible.  The effectiveness of the oven for ion trap experiments is demonstrated by measuring the trapping rates for several oven temperatures. When the oven is depleted of radium that has built up over several days, the continual decay of thorium generates a sufficient radium flux for ion trapping.

\section{oven design}

\begin{figure}
    \centering
    \includegraphics[]{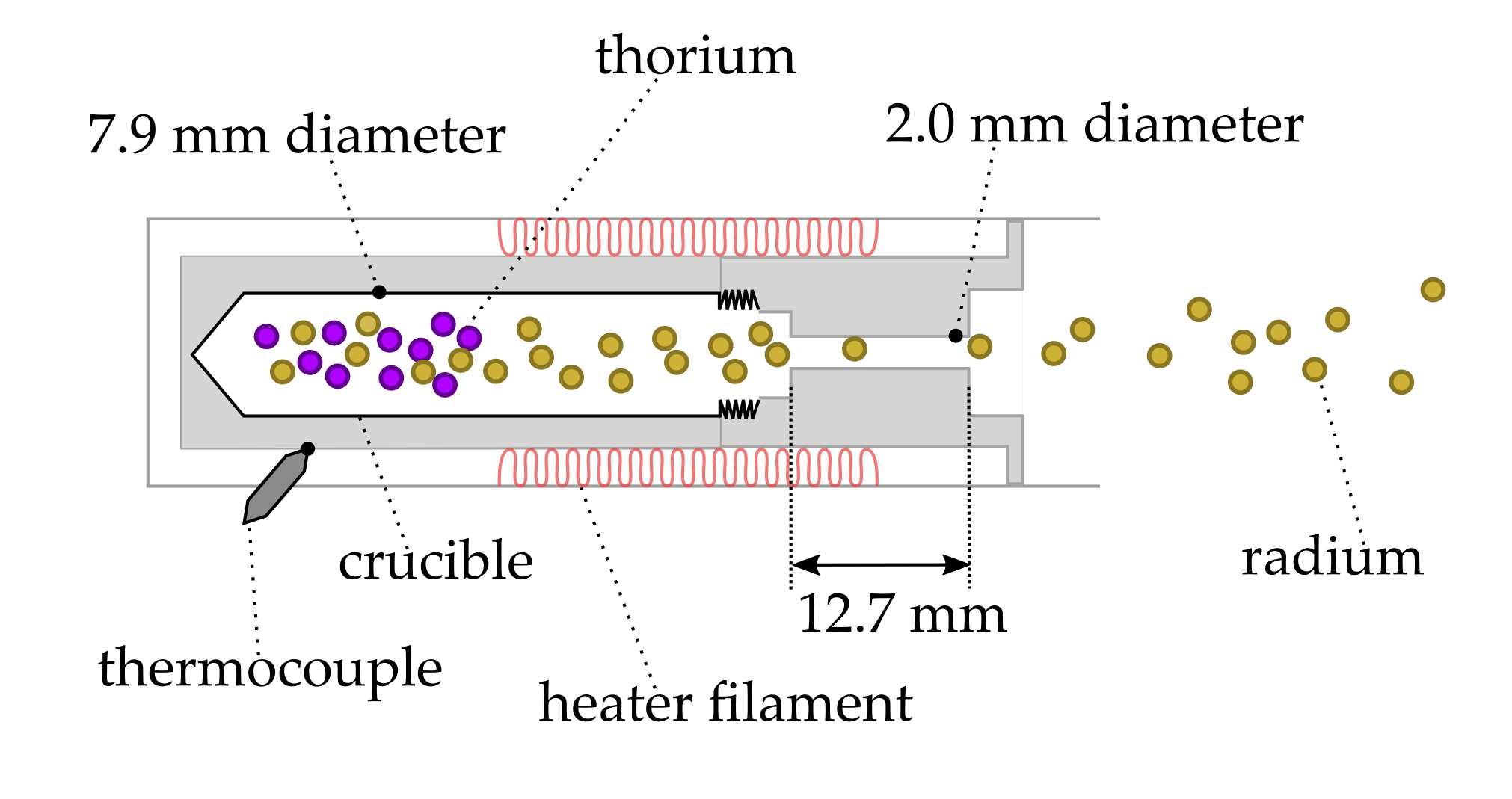}
    \caption{A schematic of the effusive oven. A titanium crucible was loaded with \iso{Th}{228} (purple circles). The crucible is resistively heated to emit a thermal beam of radium atoms (gold circles). The crucible's temperature is measured with a thermocouple in contact with its outer surface.}
    \label{fig:oven-schematic}
\end{figure}

Effusive atomic ovens are a common means to generate atomic beams for laser cooling and trapping of both neutral and ionized atoms~\cite{Ross1995,Dammalapati2009}. The oven reported here is based on an effusion cell design commonly used for molecular beam epitaxy (MBE) \footnote{The specific MBE oven used in this work was E-Science Model No. EC-010-275-0800-HL}. The effusion cell has heater wires that can heat a titanium crucible up to 1470~K. The crucible's interior is a 59-mm-long cylinder with a 7-mm diameter. The crucible cap has a 12.7-mm-long, 2-mm-diameter aperture; see Fig.~\ref{fig:oven-schematic}. We transferred 40(20)~$\mu$Ci of \iso{Th}{228}(NO$_3$)$_4$ in 0.1~M HNO$_3$ into the crucible and dried the solution in the oven by heating the crucible with a hot plate to $\sim$350~K. Once dried, we added $\sim$1~mg of strontium, attached the cap, put the crucible in the effusion cell, and installed it in a vacuum chamber.   Initially, a strontium beam from the heated oven was used for laser alignment and ion trap testing.  This strontium may play a role in reducing radium compounds which might form due to reactions with contaminants.

\section{laser cooling and trapping}

\begin{figure}
    \centering
    \includegraphics{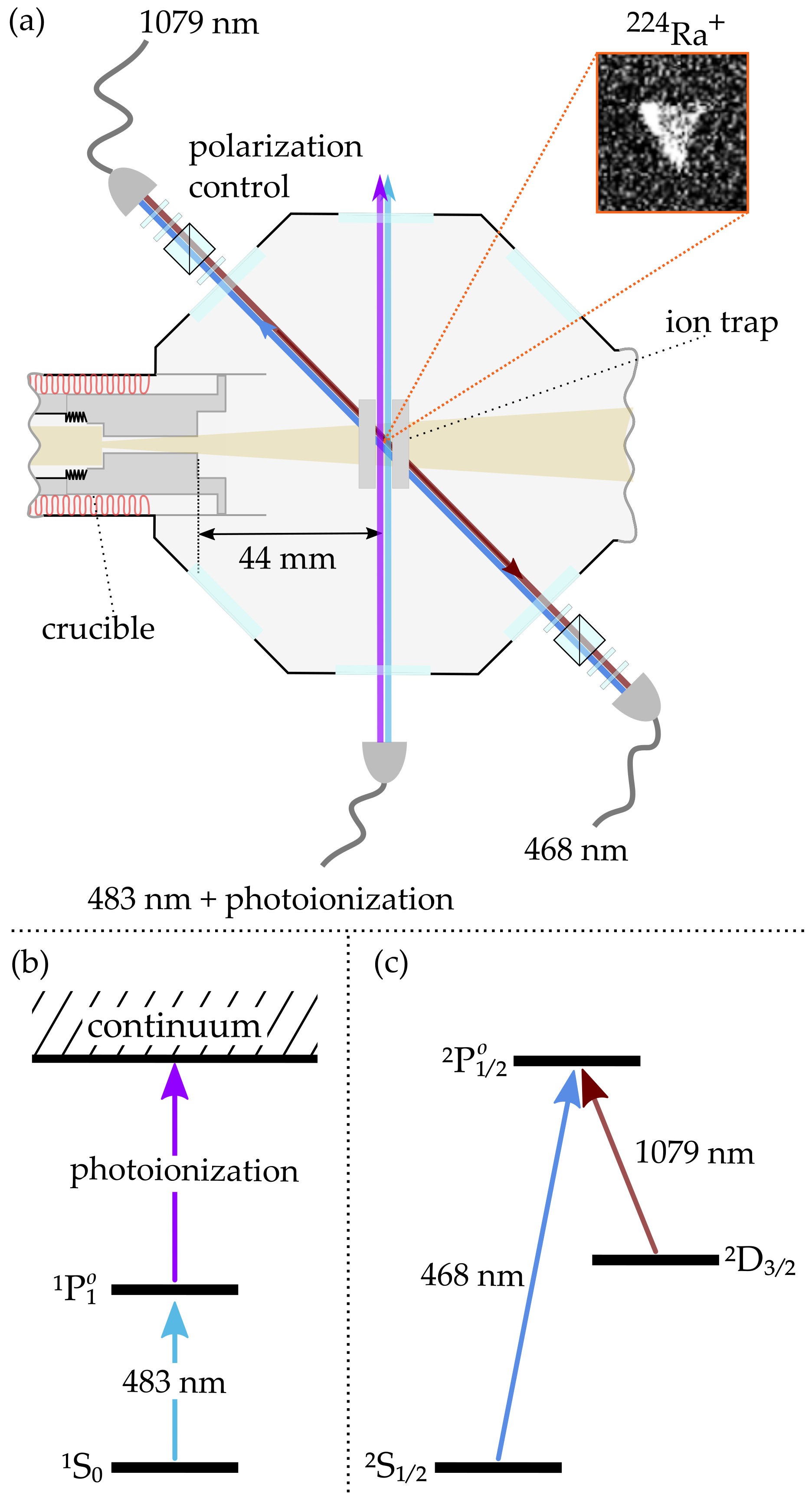}
    \caption{\textbf{(a)} A schematic of the apparatus for photoionization and laser cooling and trapping of \iso{Ra}{224}$^+$. The trap is a linear Paul trap depicted by two rf~rods. The effusive oven geometry is given in Fig.~\ref{fig:oven-schematic}. 
    \textbf{(b)} The transitions used for photoionization of \iso{Ra}{224}; 405-, 422-, and 450-nm lasers were tested on the $^1$P$_1$ transition to the continuum.  
    \textbf{(c)} The \iso{Ra}{224}$^+$ level structure with the laser cooling transitions highlighted.}
    \label{fig:pi-trap-schematic}
\end{figure}

Photoionization (PI) and subsequent laser cooling and trapping of \iso{Ra}{224}$^+$ was realized with the trap depicted in Fig.~\ref{fig:pi-trap-schematic}(a). The trap is a linear Paul trap, described in Ref.~\cite{Fan2019}; the diagonal radio-frequency (rf) electrodes are separated by 6~mm, and the end cap electrodes are separated by 15~mm. The trap center is 44~mm from the oven aperture, and the rf drive frequency is 990~kHz. Permanent magnets generate a 5.0(5)~G static magnetic field at approximately 90$^\circ$\ with respect to the laser cooling beams.

Radium atoms from the oven are photoionized in a two-stage process; see Fig.~\ref{fig:pi-trap-schematic}(b). This process is similar to the scheme used for other alkaline earth atoms~\cite{Lucas2004,Vant2006}. First, neutral radium is excited on the \mbox{$^1$S$_0 \rightarrow ^1$P$_1$} transition with 1.1~mW of 483-nm light. A photoionizing beam then drives the population from the $^1$P$_1$ level to the continuum with 1~mW of 450-nm light. The photoionizing beam waist is at the trap center and is approximately 150~$\mu$m. Laser cooling is realized with a 468-nm cooling laser red detuned from the \mbox{7$s\ ^2$S$_{1/2} \rightarrow 7p\ ^2$P$_{1/2}$} transition and a 1079-nm repump laser that drives the \mbox{6$d\ ^2$D$_{3/2} \rightarrow 7p\ ^2$P$_{1/2}$} transition; see Fig.~\ref{fig:pi-trap-schematic}(c). Scattered 468-nm light is collected by the imaging system and focused onto a photomultiplier tube (PMT) and a camera~\cite{Fan2019}.

Three PI laser wavelengths (405-, 422-, and 450-nm) were tested for the second photoionization step. These three wavelengths are all above the 458-nm ionization energy threshold from the $^1$P$_1$ state. The number of \iso{Ra}{224} atoms photoionized and detected in the trap per minute (the ion capture rate) was measured for each wavelength. The three PI wavelengths produced comparable ion capture rates for similar powers. In practice, we used 450-nm light because it had the most available power ($\sim$1~mW at the trap).

Rydberg autoionization was explored using 461- and 468-nm lasers as the second PI stage to excite Rydberg states. The PI rate with 461-nm light was lower than that with 450-nm light, and no ions were trapped when using 468-nm light. We verified that radium could still be ionized and trapped with the 450-nm PI light before, during, and after the Rydberg tests.

\begin{figure}
    \centering
    \includegraphics[]{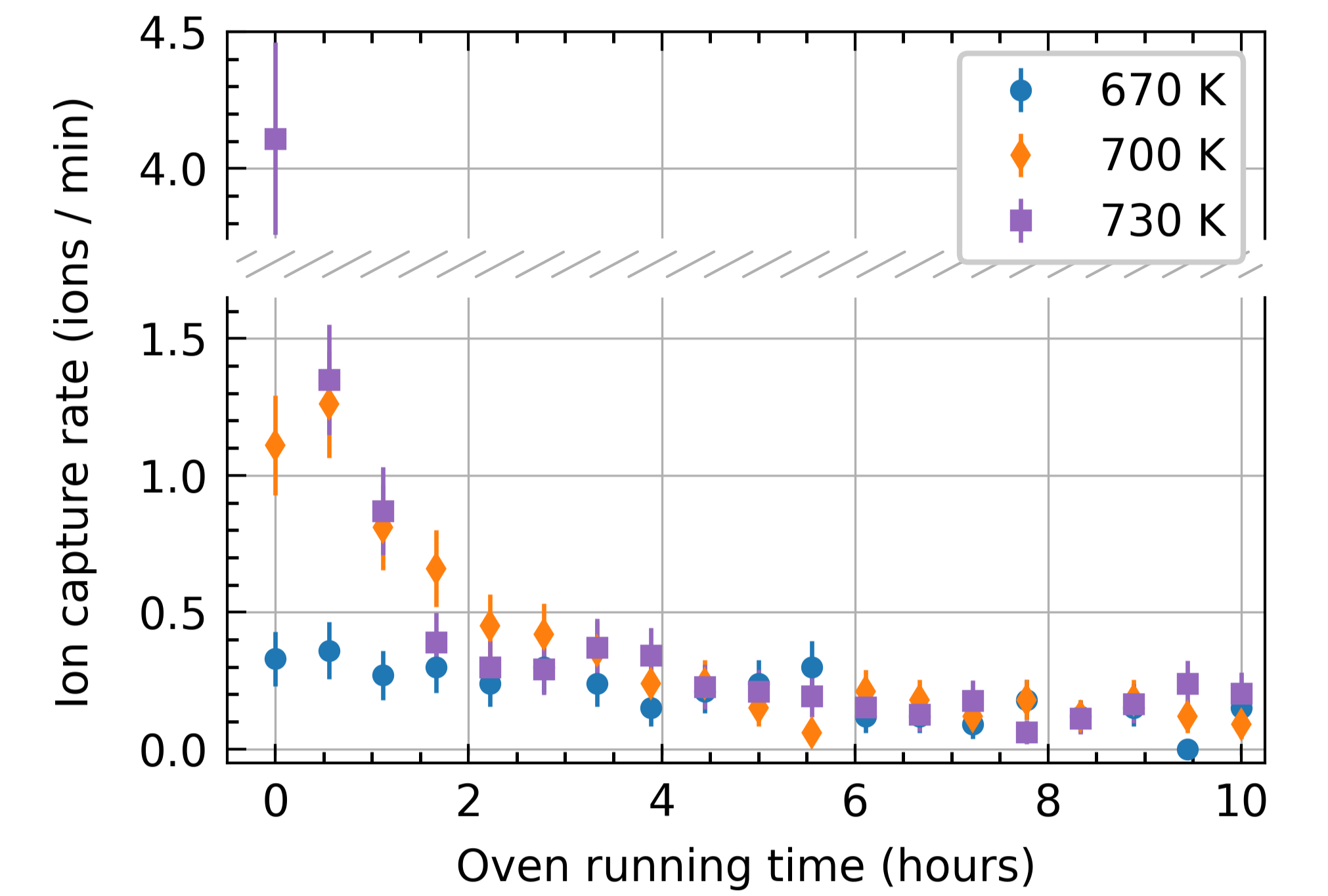}
    \caption{\iso{Ra}{224}$^+$ ion capture rates at different oven temperatures as a function of time. The 450-nm photoionization laser has 1~mW optical power and a 150-$\mu$m beam waist at the ion trap.}
    \label{fig:loading-rate-temps}
\end{figure}

We characterized the reliability and longevity of the oven source by measuring the ion capture rate for extended periods of operation. The ion capture rate was determined by an automated process which monitored PMT counts from laser cooled radium ions. When the PMT counts exceeded a detection threshold, the trap's rf power was switched off to dump the ion and then turned back on to trap the next ion. The time between turning on the rf and loading a new ion was recorded to determine the ion capture rate. The photoionization beams were applied continuously. The ion capture rate at several oven temperatures as a function of time is shown in Fig.~\ref{fig:loading-rate-temps}. The high initial capture rate, a consequence of the flux of radium atoms that have built up prior to turning on the oven, increases rapidly with oven temperature. On the order of $10^{11}$ radium atoms accumulate in one \iso{Ra}{224} half-life~\cite{radioactivedecay}. After the initial surge, the continual decay of thorium is sufficient to maintain a flux of radium atoms for trapping. A steady-state capture rate of $\sim$0.13(1)~ions/min is reached after approximately 3~h for each oven operating temperature that was tested. At higher temperatures, the steady state capture rate increases due to a combination of increased rates of radium desorption from surfaces and effusion out of the crucible's titanium walls~\cite{Beyer2003,Melconian2005}.

\section{neutral Ra\textsuperscript{224} spectroscopy}

The \mbox{$7s^2\ ^1$S$_0 \rightarrow 7s7p\ ^1$P$_1$} (\mbox{$^1$S$_0 \rightarrow ^1$P$_1$}) radium transition at 483~nm is useful for photoionization loading of radium into ion traps. The \mbox{$^1$S$_0 \rightarrow ^1$P$_1$} transition was first measured for \iso{Ra}{226} by Rasmussen in 1933~\cite{Rasmussen1933}. Subsequent spectroscopy of this transition has been performed for \iso{Ra}{226} and \iso{Ra}{225}~\cite{Trimble2009,Santra2014}. We measured the \mbox{$^1$S$_0 \rightarrow ^1$P$_1$} transition of the \iso{Ra}{224} frequency and compared our value with the previous results using the isotope shift measurements of Wendt \emph{et al.}~\cite{Wendt1987}. Saturated absorption spectroscopy of molecular tellurium (\iso{Te}{130}$_2$) was used as a frequency reference for the \iso{Ra}{224} spectroscopy~\cite{Santra2014}.

For saturated absorption spectroscopy, two parallel 483-nm probe beams and a counterpropagating pump beam which overlaps with one of the probe beams are passed through a cell containing \iso{Te}{130}$_2$ at 870(20)~K. All three beams are derived from a single laser. The probes, each 60(10)~$\mu$W, are collected on a balanced photodiode, and their signals are subtracted. The pump beam, shifted 80~MHz by an acousto-optic modulator (AOM) relative to the probe light, has 1.3(1)~mW of power and a 0.8(1)-mm beam waist at the center of the \iso{Te}{130}$_2$ cell. The AOM serves as a shutter for the pump beam, turning it on and off at a modulation frequency of 10~kHz. The same 10-kHz modulation frequency is mixed with the split photodiode output with a lock-in amplifier to measure the Te$_2$ spectrum. 

\begin{figure}[h]
    \centering
    \includegraphics[]{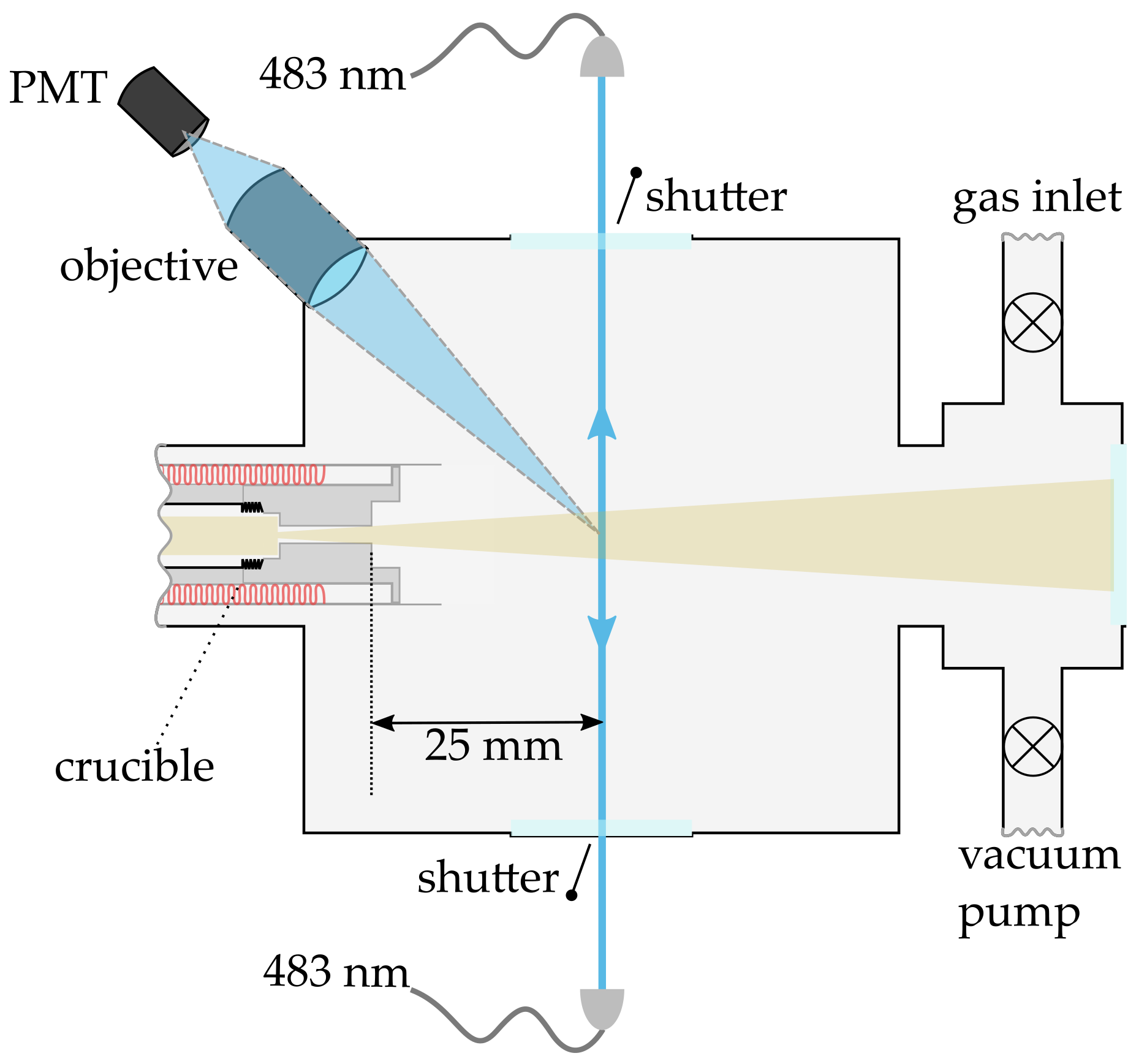}
    \caption{A schematic of the vacuum apparatus used for neutral spectroscopy of \iso{Ra}{224}. The setup uses an effusive oven with the same geometry as in Fig.~\ref{fig:oven-schematic}. A gas inlet valve allows for the introduction of gas, such as argon, into the chamber. A pair of counterpropagating 483-nm beams are perpendicular to the atomic beam. A fraction of the fluorescence from atoms excited on the \mbox{$^1$S$_{0} \rightarrow ^1$P$_1$} transition is collected by an imaging objective and focused onto a PMT.}
    \label{fig:neutral-spec-apparatus}
\end{figure}

Spectroscopy of \iso{Ra}{224} was performed in the vacuum apparatus depicted in Fig.~\ref{fig:neutral-spec-apparatus}. A thermal beam of \iso{Ra}{224} atoms is generated by heating the oven to 880(10)~K. Two counterpropagating 483-nm laser beams [each 1.5(1)~mW] are perpendicular to the radium beam 25~mm from the oven aperture. The beam waists are 3.4(1)~mm. The geometry is chosen to minimize the effect of Doppler broadening. Radium atoms are excited on the \mbox{$^1$S$_0 \rightarrow ^1$P$_1$} transition, and scattered light is collected by an imaging objective onto a PMT.

\begin{figure}[h]
    \centering
    \includegraphics[]{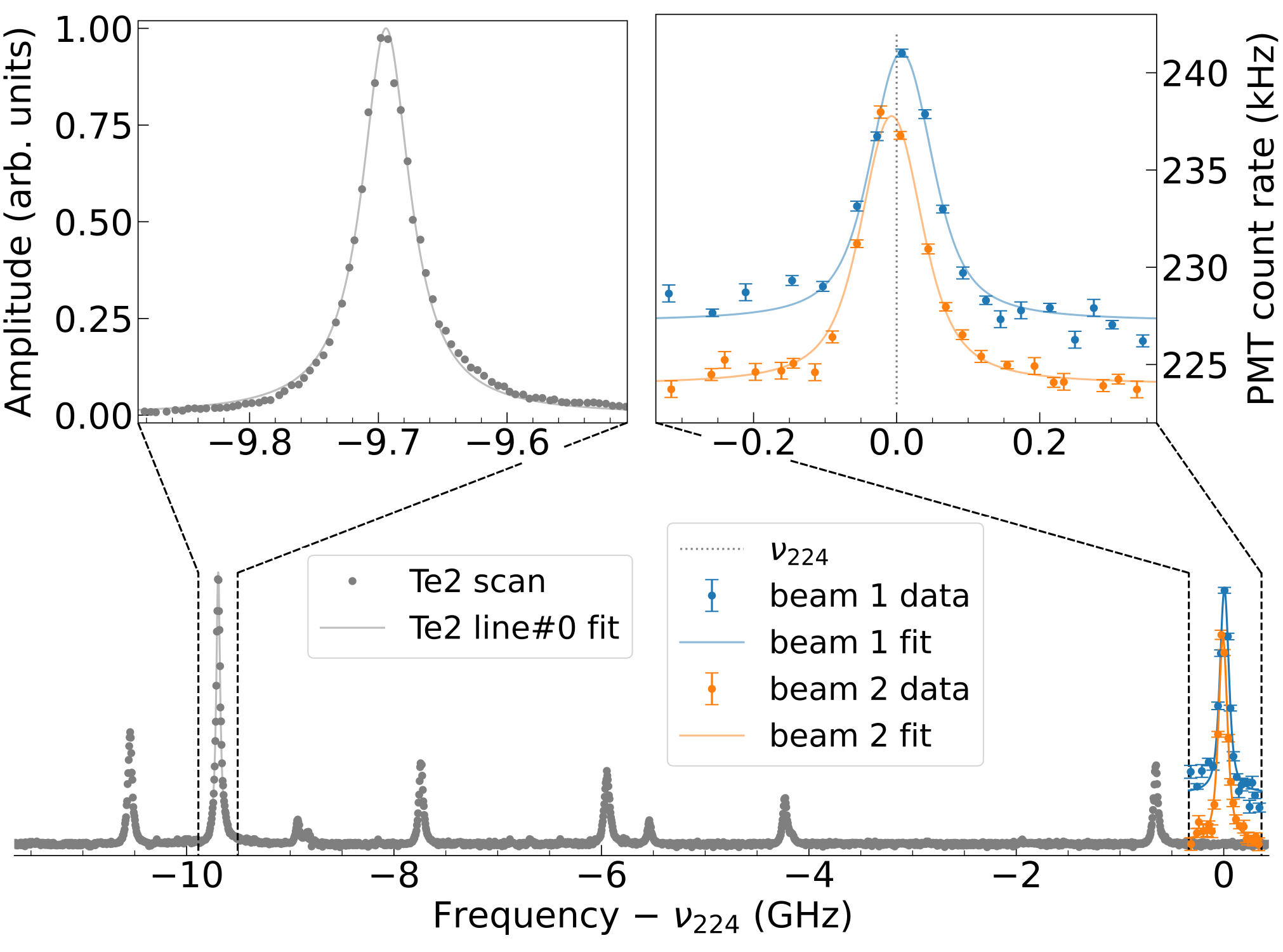}
    \caption{Spectroscopy of the \iso{Ra}{224} \mbox{$7s^2\ ^1$S$_0 \rightarrow 7s7p\ ^1$P$_1$} transition and saturated absorption spectroscopy of \iso{Te}{130}$_2$ peaks covering line -1 to line 6 as labeled in Ref.~\cite{Santra2014}, where $\nu_{224}$ is our measured \iso{Ra}{224} transition frequency. The amplitude of the \iso{Te}{130}$_2$ saturated absorption signal is normalized to $\mathrm{Te_{2}}$ line 0. The measured $\mathrm{^{224}Ra}$ \mbox{$^1$S$_0 \rightarrow ^1$P$_1$}  transition frequency is calibrated from the $\mathrm{Te_{2}}$ spectrum. The difference in peak height between beam 1 (blue) and beam 2 (orange) is due to the decay in atomic flux during the measurement.
    The left inset shows the Lorentzian fit of Doppler-free $\mathrm{Te_{2}}$ line 0.
    The right inset shows the Voigt fit of the fluorescence from the thermal \iso{Ra}{224} beam.} 
    \label{fig:483-spec}
\end{figure}

The 483~nm laser frequency is scanned continuously while the two counterpropagating beams are alternately shuttered and the two spectra are recorded. The reported frequency of the \iso{Ra}{224} \mbox{$^1$S$_{0} \rightarrow ^1$P$_1$} transition is the average frequency of both spectra using a wavemeter (10~MHz resolution) \footnote{HighFinesse WS8 (330-1180 nm)} calibrated with \iso{Te}{130}$_2$ reference line 0; see Fig.~\ref{fig:483-spec}. The fit uncertainties for the \iso{Ra}{224} and \iso{Te}{130}$_2$ transitions are their fitted half-width-at-half-maximum values, which account for model imperfections. Wavemeter drift is determined from the difference between the \iso{Te}{130}$_2$ saturated absorption spectrum measured before and after the \iso{Ra}{224} spectroscopy. The observed wavemeter drift within the measurement time of $\sim$2~h is 2~MHz. The latter \iso{Te}{130}$_2$ spectrum and the \iso{Ra}{224} spectra are shown in Fig.~\ref{fig:483-spec}. Imperfect alignment of the 483-nm beams results in a 1(9)~MHz first-order Doppler shift. The reported \mbox{$^1$S$_{0} \rightarrow ^1$P$_1$} transition frequency for \iso{Ra}{224} is \SI{621043830}{}$\pm60$ MHz.

\begin{figure}
    \centering
    \includegraphics{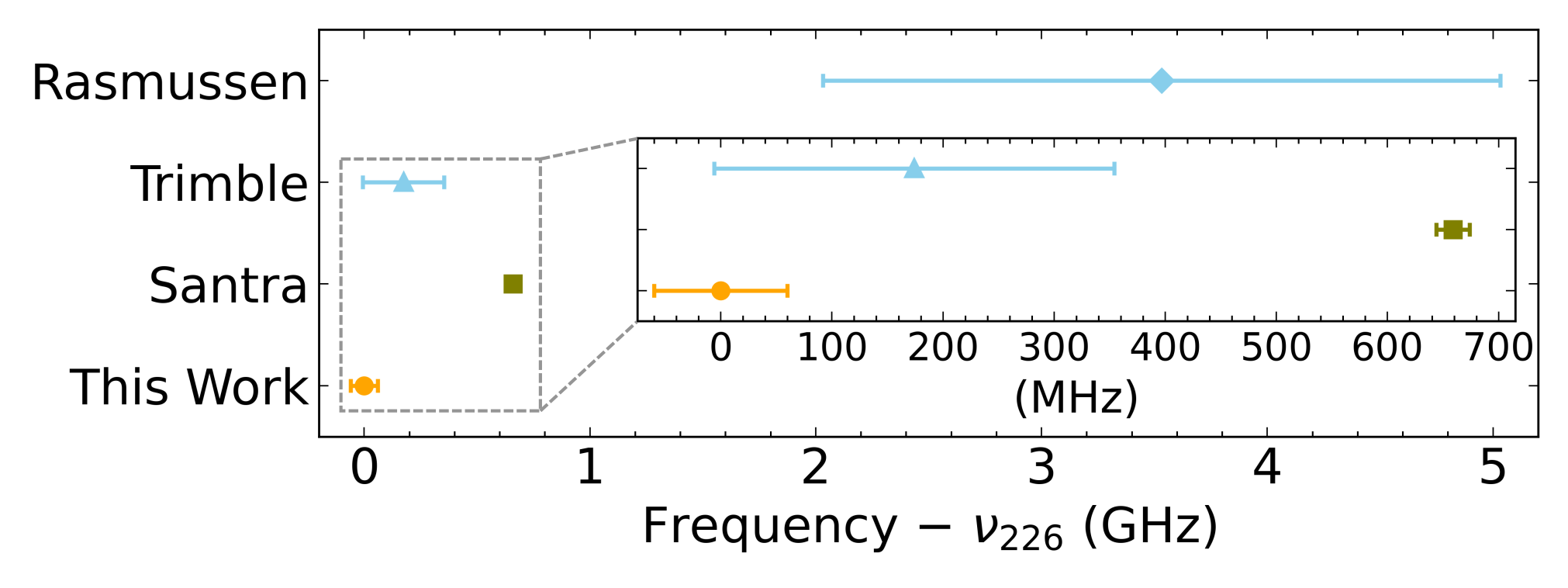}
    \caption{A comparison of the reported \iso{Ra}{226} \mbox{$7s^2\ ^1$S$_0 \rightarrow 7s7p\ ^1$P$_1$} transition frequencies where $\nu_{226}$ is our value. Frequencies reported by Rasmussen~\cite{Rasmussen1933} and Trimble \emph{et al.}~\cite{Trimble2009} are direct measurements. The frequency reported by Santra \emph{et al.}~\cite{Santra2014} is an isotope-shifted value from their measurement of the \iso{Ra}{225} transition frequency.}
    \label{fig:Ra226-freqs}
\end{figure}

We determine the \iso{Ra}{226} \mbox{$^1$S$_{0} \rightarrow ^1$P$_1$} transition frequency, \SI{621037830}{}$\pm60$ MHz, by the isotope shifts of \iso{Ra}{224} and \iso{Ra}{226} with respect to \iso{Ra}{214}~\cite{Wendt1987}. There is a 660-MHz discrepancy between our value for the \mbox{$^1$S$_{0} \rightarrow ^1$P$_1$} transition frequency and the value reported in Ref.~\cite{Dammalapati2016}; see Fig.~\ref{fig:Ra226-freqs}.

\section{null results}\label{sec:null-tests}

Short-lived radioisotopes are challenging to work with, particularly due to the difficulty of producing a sufficient atom flux for neutral spectroscopy and ion trapping. Radium poses further difficulties as it is reactive. Different techniques were tested in the neutral spectroscopy setup of Fig.~\ref{fig:neutral-spec-apparatus} with indeterminate results or uncertain effectiveness, some of which are described here as paths for future exploration. 

Argon gas with a pressure of $\sim$100~Torr was flowed through the neutral spectroscopy apparatus in an effort to slow the radium atoms after nuclear decay and prevent them from becoming deeply buried in the titanium walls of the crucible~\cite{Wense2016}. No increase in PMT counts on the \mbox{$^1$S$_0 \rightarrow ^1$P$_1$} transition was observed.

We also investigated reducing agents. When \iso{Th}{228} was loaded into the crucible without any reducing agents, no \mbox{$^1$S$_0 \rightarrow ^1$P$_1$} transition peak was observed. After adding in $\sim$1~mg of strontium, 200~mg Zr powder, and 50~mg BaCO$_3$ to the crucible to reduce radium compounds \cite{Santra2014}, there was an increase in the neutral radium spectroscopy signal compared with when only strontium was used. However, it is unclear if the lack of signal without reducing agents was due to a low pressure of reactive background gas molecules rather than the lack of reducing agents. The pressure of the neutral spectroscopy chamber [$\sim$10$^{-5}$~Torr with an oven temperature of 880(10)~K] was three orders of magnitude higher than that of the ion-trapping apparatus. At the lower pressures achieved in the ion trap, reducing agents may not be necessary.

\section{conclusion}

This work lowers the barrier to using the short-lived \iso{Ra}{224} isotope in cold-atom experiments.  An effusive oven based on the decay of thorium is a reliable source of radium atoms for ion-trapping experiments and could be used for cold-neutral-atom experiments depending on atom number requirements and acceptable activity.  The source efficiency may be increased with more advanced oven nozzle geometries~\cite{Senaratne2015}.   An oven based on the same principle may be used to laser cool and trap \iso{Ra}{225} ions, produced via the decay of \iso{Th}{229} (7825-yr half-life~\cite{Essex2018}), or \iso{Ra}{226} ions, produced via the decay of \iso{Th}{230} (\SI{75400}{}-yr half-life). Such ovens are robust to radium depletion, e.g., due to overheating the oven, as radium is continuously repopulated by thorium.   

\section*{acknowledgements}

We thank Chris Palmstr\o m, Dave DeMille, Bodhaditya Santra, and Lorenz Willmann for helpful discussions, Craig Bichrt for help with the MBE oven configuration, and Dave Patterson for lending a 461-nm laser. We thank Chris Greene and Miguel A. Alarcón for guiding calculations. M.F. and H.L. were supported by DOE Award No. DE-SC0022034. S.K. was supported by ONR Grant No. N00014-21-1-2597. R.A.R., R.K., C.A.H., M.S.L., and A.M.J. were supported by NSF Award No. PHY-2146555,  NIST Award No. 60NANB21D185, the Heising-Simons Foundation, the W.M. Keck Foundation, and the Eddleman Center. A.N.G. acknowledges the support of startup funds and Michigan State University. The isotope used in this research was supplied by the U.S. Department of Energy Isotope Program, managed by the Office of Isotope R\&D and Production.

\bibliography{references}

\end{document}